\documentclass[preprint2]{aastex62}
\usepackage{natbib}
\usepackage{color} % {\color{red} This text is red.}
\usepackage{aas_macros} %automatic handling of bibTeX entries from ADS
\usepackage{ulem}
\received{July 2018}
\revised{}
\accepted{}
\begin{document}

\title{Anisotropy of the Milky Way's stellar halo
  using K giants from LAMOST and {\it Gaia}}

%\correspondingauthor{Sarah A. Bird}
\email{sarahbird@shao.ac.cn, xuexx@nao.cas.cn, liuchao@nao.cas.cn, jshen@shao.ac.cn, cflynn@swin.edu.au}

\author{Sarah A. Bird} \affil{Shanghai Astronomical Observatory, 80
  Nandan Road, Shanghai 200030, China}
\affil{Key Laboratory of Optical Astronomy, National Astronomical
  Observatories, Chinese Academy of Sciences, Beijing 100012, China}

\author{Xiang-Xiang Xue$^{2}$} 
\author{Chao Liu$^{2}$}

\author{Juntai Shen$^{1}$}
\affil{College of Astronomy and Space Sciences, University of
  Chinese Academy of Sciences, 19A Yuquan Road, Beijing 100049, China}

\author{Chris Flynn}
\affil{Centre for Astrophysics and Supercomputing, Swinburne
  University of Technology, Post Office Box 218, Hawthorn, VIC 3122,
  Australia.}

\author{Chengqun Yang$^{2,}$}
\affil{School of Astronomy and Space Science, University of Chinese Academy of Sciences, Beijing 100049, China}

\begin{abstract}

The anisotropy parameter $\beta$ characterizes the extent to which
orbits in stellar systems are predominantly radial or tangential, and
is likely to constrain, for the stellar halo of the Milky Way,
scenarios for its formation and evolution. We have measured
 $\beta$ as a function of Galactocentric radius from $5-100$
kpc for 7664
metal poor ([Fe/H] $<-1.3$) halo K giants from the
LAMOST catalog with line-of-sight velocities and distances, matched to
proper motions from the second {\it Gaia} data release. We construct full
6-D positions and velocities for the K giants to directly measure the
three components of the velocity dispersion $(\sigma_r, \sigma_\theta,
\sigma_\phi)$ (in spherical coordinates). We find that the orbits in
the halo are radial over our full Galactocentric distance range
reaching over 100 kpc.  The anisotropy remains remarkably unchanged with Galactocentric radius
from approximately 5 to 25 kpc, with an amplitude that depends on the
metallicity of the stars, dropping from $\beta \approx 0.9$ for $-1.8 \leq$ [Fe/H]
$< -1.3$ (for the bulk of the stars) to $\beta \approx 0.6$ for the lowest
metallicities ([Fe/H] $< -1.8$).
Considering our sample as a whole, $\beta\approx0.8$ and, beyond 25 kpc, the orbits gradually become
less radial and anisotropy decreases to $\beta<0.3$ past 100
kpc. Within 8 kpc, $\beta<0.8$. The measurement of anisotropy is
affected by substructure and streams, particularly beyond a
Galactocentric distance of approximately 25 kpc, where the Sagittarius
stream is prominent in the data. These results are complimentary to recent 
analysis of simulations by Loebman et al. and of SDSS/{\it Gaia} DR1 data by Belokurov et al.

\end{abstract}


\section{Introduction} \label{sec:intro}

The Milky Way's halo has long provided constraints on scenarios for
the formation and evolution of our galaxy, as it contains relics from
the earliest stages of the process. In the dominant theoretical
framework of galaxy formation, in which small scale structures
condense most quickly under gravity in $\Lambda$CDM, 
Milky Way-like halos form early
out of large numbers of minor and some major mergers, and have
progressed well around the present day sized-Galaxy
before the bulge and
disk formation begins.

A fundamental constraint is provided by the stellar orbital families
that develop during the assembly. 
The velocity
anisotropy parameter $\beta$ \citep{Binney1980,Binney2008} 
characterizes the extent to which orbits
in halos are predominantly radial or tangential and is defined
through the velocity dispersions $\sigma$ of the stars. Throughout this paper we use spherical
coordinates $(r,\theta,\phi)$ as described in \citet[Appendix B]{Binney2008}\footnote{Different coordinate systems (e.g. Cartesian, cylindrical, spherical), symbols (e.g. $\theta$, $\phi$), and schemes (e.g. (radial, polar, azimuthal), (radial, azimuthal, polar)) are used throughout the literature, which can be a source of possible confusion. 
}. In this system, the anisotropy parameter is defined as
\begin{equation}
\beta = 1 - (\sigma_\theta^2+\sigma_\phi^2)/(2\sigma_r^2),
\end{equation}
where $r=r_\mathrm{gc}$ is the distance from the Galactic Center, $\theta$ is the polar angle $0<\theta<\pi$ starting from the positive $Z$ axis and increasing towards the $X-Y$ plane, and $\phi$ is the azimuthal angle $-2\pi<\phi<2\pi$ for which $\phi = 0$ on the positive $X$ axis and $\phi$ increasing towards
    the positive $Y$ axis.
In an isothermal system, $\sigma_\theta = \sigma_\phi = \sigma_r$, and
$\beta = 0$. For stellar systems in which the orbits are
predominantly radial, $\beta > 0$. For predominantly tangential orbits,
$\beta < 0$. By definition, $1 > \beta > -\infty$.

\citet{Hattori2017} and \citet{Loebman2018} have thoroughly reviewed
the history of $\beta$ measurement using halo stars within a few kpc
of the Sun, programs to measure $\beta$ from line-of-sight velocities
out to tens of kpc from the Sun, and of the behavior of $\beta$ in
$N$-body and hydrodynamical simulations of Milky Way-like
galaxies.

Until the advent of {\it Gaia}, large stellar samples with direct
measurements of all three components of $\sigma$ were limited to
within the neighborhood of the Sun, i.e. within a few kpc. Locally,
studies with e.g. RR Lyrae, K-giant, blue horizontal branch
(BHB), and subdwarf stars find $\beta \approx 0.5 $-$0.7$, and the
orbits are thus predominantly radial. 
For example\footnote{We quote the results directly from the paper and in parentheses we rewrite the results into the spherical ($r$,$\theta$,$\phi$) system used in the current paper.},
\citet{Morrison1990} derive 
$(\sigma_r, \sigma_\phi, \sigma_\theta) = (133 \pm 8, 98 \pm 13, 94 \pm 6)$ km s$^{-1}$ for halo G and K giants up to
a few kpc distant 
($(\sigma_r, \sigma_\theta, \sigma_\phi) = (133 \pm 8, 94 \pm 6, 98 \pm 13)$ km s$^{-1}$ for which $\beta \approx 0.5$). \citet{Chiba1998}
estimate $\beta = 0.52 \pm 0.07$ for a sample of halo stars ([Fe/H] 
$< -1.6$) within a few kpc of the Sun, based on the measurement of
$(\sigma_U, \sigma_V, \sigma_W) = (161 \pm 10, 115 \pm 7, 108 \pm 7)$ km s$^{-1}$
($(\sigma_r, \sigma_\theta, \sigma_\phi) \approx (161 \pm 10, 108 \pm 7, 115 \pm 7)$ km s$^{-1}$). 
\citet{Smith2009} find $\beta = 0.69 \pm 0.01$ using
$\approx 1700$ halo subdwarfs in SDSS Stripe 82 data up to 5 kpc
distant, for which $(\sigma_r, \sigma_\phi, \sigma_\theta) = (143 \pm 2, 82 \pm 2, 77 \pm 2)$ km s$^{-1}$
($(\sigma_r, \sigma_\theta, \sigma_\phi) = (143 \pm 2, 77 \pm 2, 82 \pm 2)$ km s$^{-1}$).

Efforts to understand whether such radial orbits are the case
elsewhere in the Galaxy --- and particularly outside the Solar Circle
--- have been hampered by the lack of proper motions. The difficulties
in accurately estimating velocity dispersions and velocity anisotropy,
due largely to distance errors and lack of proper motions, are
explored by, e.g., \citet{Schonrich2011,Beers2012,Fermani2013}, and
\citet{Hattori2017}.
Out to a few tens of kpc, aided by models of the mass
distribution in the Galaxy, line-of-sight velocities have been used to
indirectly determine $\beta$ from a range of tracers, 
including BHB stars, K giants, and F-type stars \citep[e.g.]{Sommer-Larsen1994,Thom2005,Kafle2012,Kafle2014,King2015}. 
Anisotropy has been measured using {\it HST} proper motion measurements for 13 main sequence turn-off halo stars with distance greater than ten kpc by \citet{Deason2013}. \citet{Cunningham2016}, as part of the HALO7D project,
combine {\it HST} proper motions with line-of-sight velocities for 13 main sequence turn-off stars to measure $\beta$ beyond ten kpc.
As
discussed in detail by \cite{Loebman2018}, these studies have differed
substantially in their results, with no clear picture of the behavior
of $\beta$ with Galactocentric radius emerging.

In this study, we use 7664 halo K giants for which we
have radial velocities and stellar physical parameters 
\citep{Wu2011,Luo2012,Wu2014} in the LAMOST data release DR5 (for
information on LAMOST and its spectroscopic survey, we refer the
reader to \citet{Cui2012}, \citet{Deng2012}, \citet{Zhao2012}, and
\citet{Luo2015}). Proper motions for the vast majority of these stars
are available in the second data release (DR2) by {\it Gaia}
\citep{GaiaCollaborationBrown2018} of April 2018. We compute $V_r,
V_\theta, V_\phi$ velocities of the halo giants as functions of
Galactocentric radius $r_\mathrm{gc}$ and metallicity
[Fe/H]. Substructure within the overall halo is clearly seen in the
velocity-distance planes, such as the prominent Sagittarius stream
\citep{Belokurov2006ApJ.642L.137}. The velocity dispersions of the
radial and tangential components of the halo are found to differ
substantially, and we present the anisotropy parameter $\beta$ as a
function of Galactocentric radius and stellar metallicity from this
sample, finding that $\beta$ is similar elsewhere in the halo as in
the Solar neighborhood, with orbits very predominantly radial. We find
nearly no dependence in $\beta$ with Galactocentric radius
$r_\mathrm{gc}$ out to approximately 25 kpc, despite the velocity
dispersions changing markedly with $r_\mathrm{gc}$. We additionally
find that the amplitude of $\beta(r)$ reduces with decreasing
metallicity, as has been found very recently for much more nearby main
sequence halo stars by \citet{Belokurov2018}. We compare our
results for $\beta$ to theoretical work using simulations of halo
formation: these studies had successfully anticipated the result that
the halo is predominantly radial, but to the best of our knowledge had
not anticipated that $\beta$ would show such metallicity dependence.

In Section \ref{lamostkgiants}, 
we present our sample of 7664 halo K giants,
extending to over 100 kpc, for which we have LAMOST velocities,
distances and metallicities, matched to {\it Gaia} DR2 proper motion
data. In Section \ref{sec:results} we present the kinematics of the sample, and show
that the orbital families are predominantly radial at all radii
probed, through the anisotropy parameter $\beta$. We find that the
amplitude of $\beta$ is a function of metallicity, substantially
reduced relative to the metal rich halo stars, but still radial. 
We discuss our results in terms of observational work in
the literature, and $N$-body/hydrodynamical simulations of the
formation and evolution of galaxy halos.
In
Section \ref{sec:conclusion}, we
draw our conclusions.

\section{LAMOST sample of halo K giants}
\label{lamostkgiants}

LAMOST has completed its fifth year of observing and has planned the
data release of the LAMOST DR5 catalog for late 2018, to which we have
acquired early access. Our halo K-giant sample has been selected using effective
    temperatures and surface gravities computed on the basis of spectral
    line features \citet{Liu2014}. The specific selection criteria are
    $4000<T_\mathrm{eff}/\mathrm{K}<4600$ with $\log g<3.5$ dex and
    $4600<T_\mathrm{eff}/\mathrm{K}<5600$ with $\log g<4$ dex. As
    shown in \citet{Liu2014}, these criteria select red-giant-branch
    stars over their full range of metallicities, and the
    contamination by stars other than K giants is small, $<2.5$\%.

\begin{figure}
    \includegraphics[width=1.00\columnwidth]{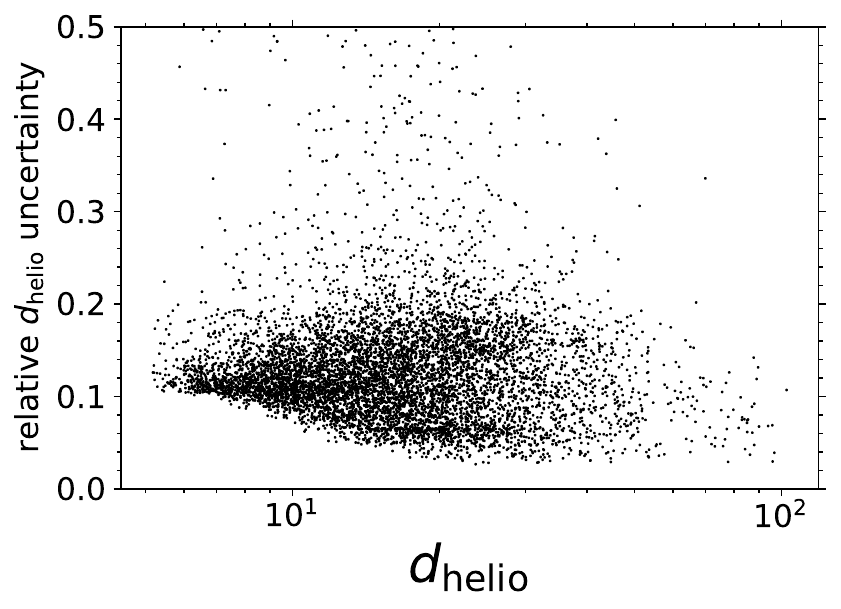}
    \caption{Relative distance uncertainties for our halo K-giant sample, shown as a function of heliocentric distance $d_\mathrm{helio}$.
}
  \label{fig:relativederr}
\end{figure}

The distances for the halo K giants are estimated using the method of
\citet{Xue2014}, which we note is the same as the one used for Sloan
Digital Sky Survey (SDSS) DR9 \citep{Ahn2012} K-giant stars in the SEGUE
project \citep{Yanny2009}. 
The photometry used in the distance estimation is from Pan-STARRS1 \citep{Chambers2016} and SDSS and we calibrated Pan-STARRS1 to the SDSS photometric system.
We use the reddening/extinction/dust maps of \citet{Schlegel1998} to estimate the extinction towards our stars. Studies show that these maps have biases which need to be corrected \citep[e.g. ][]{Cambresy2005}. Our sample stars are all more than 5 kpc above the Galactic disk, so the extinctions derived from the \citet{Schlegel1998} maps, which are dominated by dust closer than 1 kpc to the Galactic plane, are appropriate. We assume the extinction at infinity for our halo star sample. 
Alternative maps are also available to be used, for example \citet{Hakkila1997}, \citet{Cambresy2005}, 
and \citet{Schlafly2011}.
We convert the \citet{Schlegel1998} $E(B-V)$ to SDSS filters using the conversion coefficients found on Table 6 of \citet{Schlafly2012} (for $R_V=A(V)/E(B-V)=3.1$, the traditional value for the diffuse interstellar medium \citep{Cardelli1989}). 
We use the same luminosity ($\propto L^{-1.8}$) 
and density ($\propto r^{-3}$) profile priors for the \citet{Xue2014} Bayesian method, and
  the LAMOST K-giant [Fe/H] distribution as the metallicity prior. 
The Bayesian method allows the uncertainties of the input values, namely apparent magnitude, color, and metallicity, to be propagated properly to the final derived distances. 
More detailed analysis is described in \citet{Xue2014} for data from SDSS and similar conclusions can also be made for our LAMOST sample of halo K giants.
The propagated uncertainties for SDSS K giants yield 
a median distance precision of 16\% and for LAMOST K giants of 13\% (see Fig. \ref{fig:relativederr}). 
The smaller median relative distance uncertainty for LAMOST K giants is due to smaller errors in Pan-STARRS1 photometry and LAMOST metallicities compared to the SDSS photometry and metallicities used for SDSS K giant distances. We note that the difference is small and thus the distances and uncertainties are comparable for LAMOST and SDSS K giants.
Using the method presented in \citet{Xue2014}, we select giant branch stars from
above the horizontal branch to the tip of the red-giant branch.
\citet{Xue2014} derived a relation between [Fe/H] and the $(g - r)_0$
color of the giant branch at the level of the horizontal giant branch,
using eight globular clusters with $ugriz$ photometry from
\citet{An2008}, with cluster data given in Table 3 of
\citet{Xue2014}. The [Fe/H] and $(g - r)^\mathrm{HB}_0$ for the
horizontal branch/red clump of the clusters follow a quadratic
polynomial, $(g - r)^\mathrm{HB}_0 = 0.087\mathrm{[Fe/H]}^2 +
0.39\mathrm{[Fe/H]} + 0.96$, 
as shown in Fig. 2 of \citet{Xue2014}.
We work with giants above the horizontal branch to avoid possible
contamination by (metal rich) red clump stars, and also in order to
probe as distantly as possible in the halo by using the brightest giants.

%--------------------------------------------------------------------------
%--------------------------------------------------------------------------

\begin{figure*}
  \begin{tabular}{cc}
\includegraphics[width=1\columnwidth]{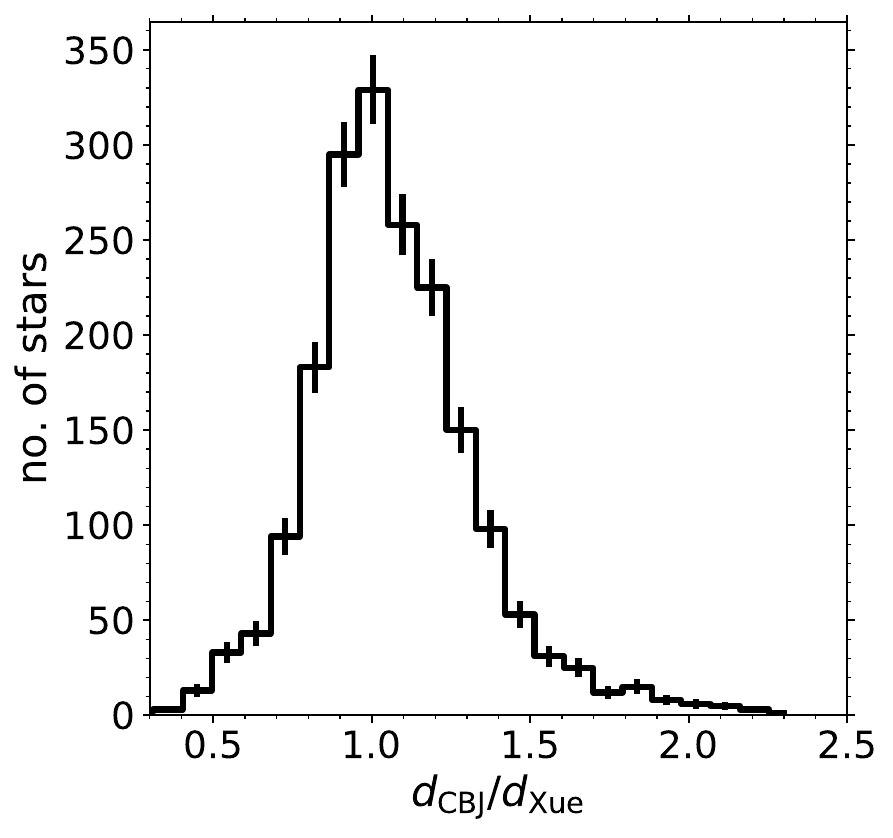}&
\includegraphics[width=1\columnwidth]{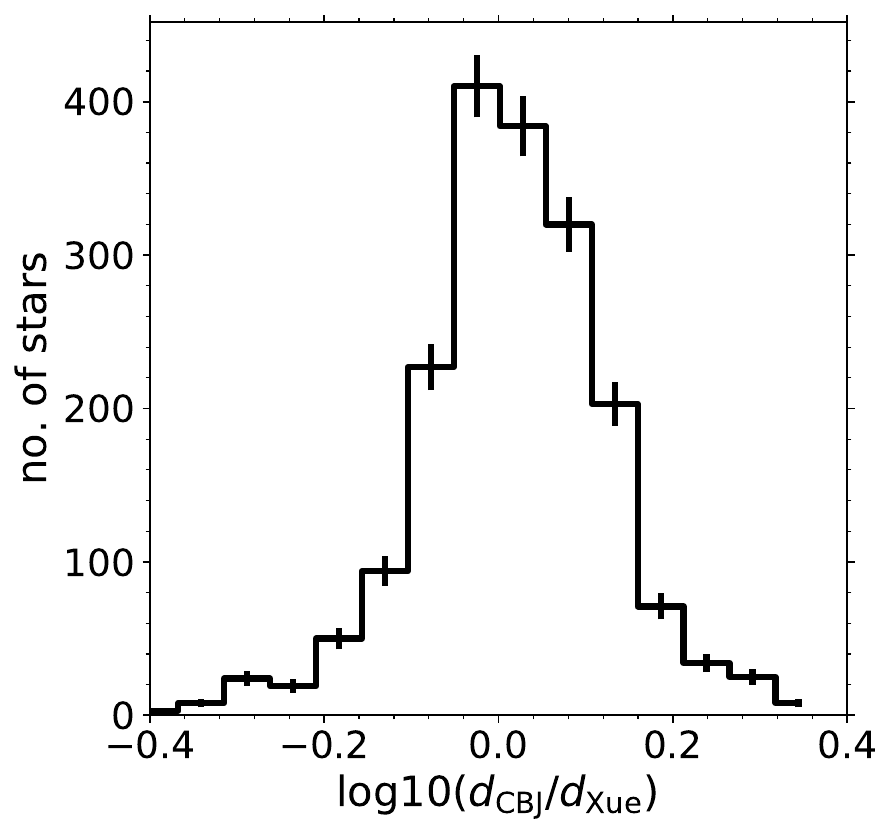}\\
\includegraphics[width=1\columnwidth]{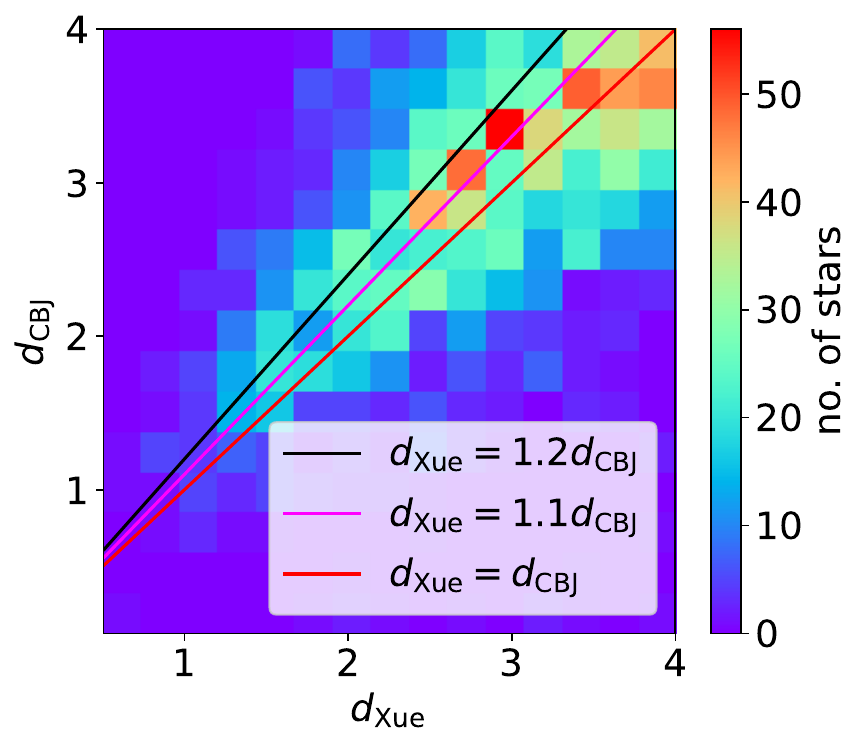}&
    \includegraphics[width=1\columnwidth]{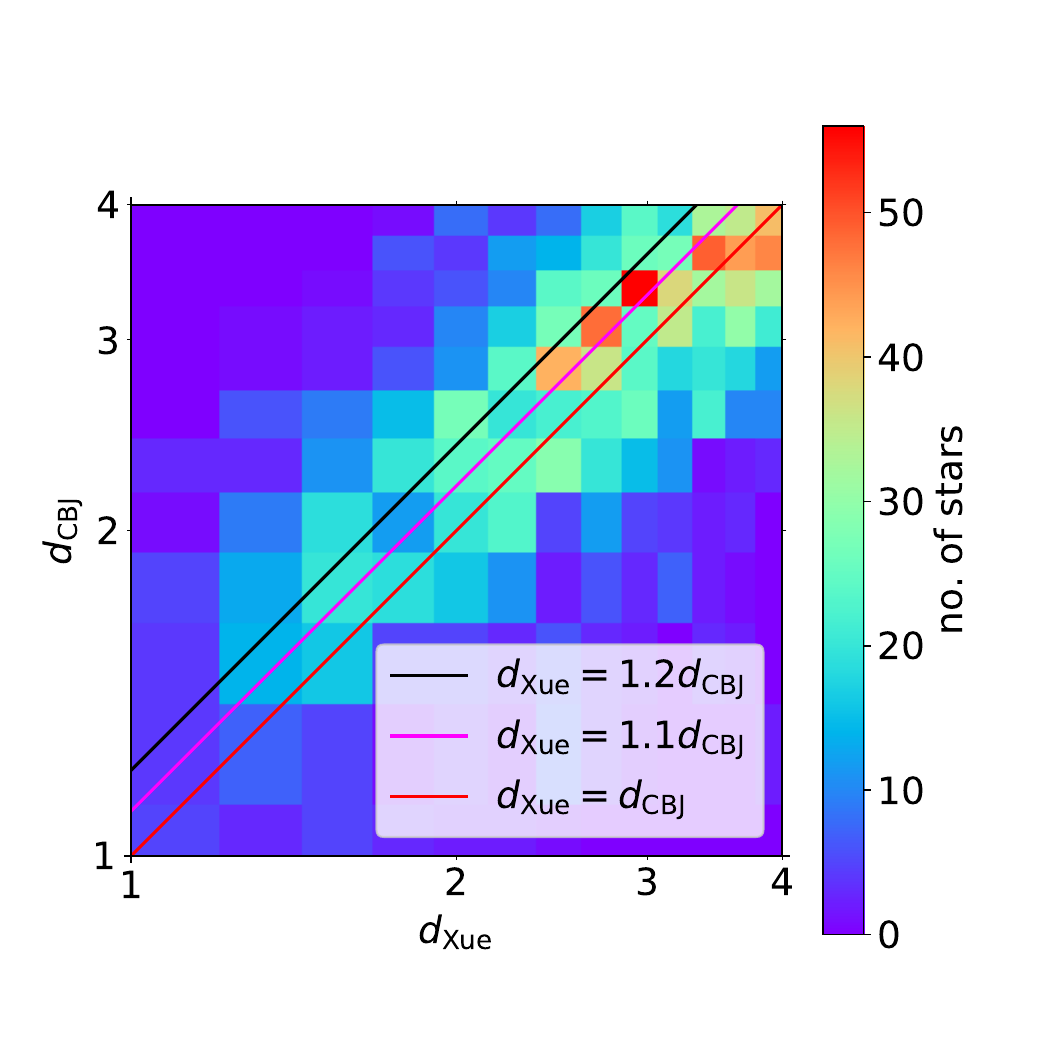}
  \end{tabular}
  \caption{Comparison of the heliocentric distances $d_\mathrm{CBJ}$ estimated by \citet{Bailer-Jones2018} to the heliocentric distances $d_\mathrm{Xue}$ of
    \citet{Xue2014}, for halo K giants ([Fe/H$<-1$) in LAMOST data and within 4 kpc of the Sun.  {\bf Upper panels:} histograms of the ratio 
$d_\mathrm{CBJ}/d_\mathrm{Xue}$ shown in linear and log scales. The histograms show a systematic difference between the two distance scales of 10\%, in the sense that
      the \citet{Xue2014} distances are closer by 10\% on average than those of \citet{Bailer-Jones2018}. {\bf Lower panels:} density map comparisons
      of the distances estimated by \citet{Bailer-Jones2018} to those of \citet{Xue2014}. Note the axes are linear in the left panel and logarithmic in the right panel.
      The lines mark the 1:1 relation between the scales, and systematic shifts of 10\% and 20\% between them.
 }
  \label{fig:distance-compare}
\end{figure*}

We have compared the \citet{Xue2014} distances to
    distances derived by \citet{Bailer-Jones2018} using {\it Gaia} DR2
    parallaxes. We select K giants
    within 4 kpc of the Sun where distances derived by parallax are
    accurate, and stars with [Fe/H]$<-1.0$ dex. In order to increase our sample size when estimating
    distances with the \citet{Xue2014} method, we use photometry from
    Pan-STARRS1, SDSS, and additionally APASS \citep[AAVSO Photometric
      All-Sky Survey,][]{Henden2014,Henden2015,Henden2016}, as this boosts the sample size. We finally
    select only stars with relative distance uncertainty better than
    40\% $(\Delta d_\mathrm{helio}/d_\mathrm{helio}<0.4$) for both methods.
    Fig. \ref{fig:distance-compare} shows the results. The top panels show histograms of the ratio of the two distance scales
    (denoted $d_\mathrm{CBJ}$ and $d_\mathrm{Xue}$ for \citet{Bailer-Jones2018} and \citet{Xue2014}, respectively).
    There is close agreement between the scales, although we find
    there is an approximately 10\% difference between the distance scales, with the \citet{Bailer-Jones2018} being greater on average
    (lower panels). This has small systematic effect on the anisotropy measurements of the halo sample, and is discussed further in
    Section \ref{sec:distance-beta}.
Adopting the \citet{Bailer-Jones2018} scale instead of the
    \citet{Xue2014} scale reduces the anisotropy parameter $\beta$ somewhat (see Section \ref{sec:distance-beta}
 and Fig. \ref{fig:xue_cbj_beta}).

%--------------------------------------------------------------------------
%--------------------------------------------------------------------------

Besides selecting K giants from LAMOST, \citet{Liu2014}
    also provide a distance calculation method for K giants.
    \citet{Liu2014} used 2MASS \citep{Cutri2003,Skrutskie2006}
    photometry and synthetic isochrones to derive distances, also
    using a Bayesian method similar to \citet{Xue2014}. We choose to
    use the method of \citet{Xue2014} for several reasons.
    Photometry from Pan-STARRS1 and SDSS, which is used by the method
    of \citet{Xue2014}, is more suitable for our K-giant sample than
    2MASS.
\citet{Xue2014} use globular cluster fiducials observed by SDSS. Due to these reasons, the \citet{Xue2014} method is more precise and we choose to use these distances.

Our sample consists of stars with [Fe/H]$<-1.3$ and which are well above and
below the Galactic disk, using the cut $|Z|>5$ kpc (where $Z$ is
height above the Galactic disk mid-plane).

The heliocentric line-of-sight LAMOST velocities need to be
corrected by adding $5.7$ km s$^{-1}$, a systematic bias which has been
calculated by comparing LAMOST velocities to SDSS/SEGUE 
\citep{Tian2015} and to {\it Gaia} \citep{Schonrich2017}.
The LAMOST giants were cross-matched to {\it Gaia} DR2 to attain
proper motions. Our total sample consists of 7664 halo K giants
with distances, line-of-sight velocities, and proper motions.

We derive Galactocentric Cartesian coordinates $(X,Y,Z)$ and velocities
$(U,V,W)$ using the conventions in {\tt astropy} \citep{astropy2018}. 
The Sun is located at $X_\odot = -8.3$ kpc and $Z_\odot = 29$ pc.
The local standard of rest (LSR) velocity is $v_\mathrm{lsr}=220$ km s$^{-1}$, and the
motion of the Sun with respect to the LSR is $(U, V, W)_\odot$ =
(11.1, 12.2, 7.3) km s$^{-1}$ \citep{Schonrich2010}.
We have tested different values for $X_\odot$ (8 kpc), $V_\mathrm{lsr}$ (190, 250 km s$^{-1}$), $V_\odot$ (1, 25 km s$^{-1}$, which are extreme values taken from \citet{Robin2017} and \citet{Bovy2015}, respectively), and $W_\odot$ (20 km s$^{-1}$)
and conclude that our main results remain unchanged.

We define spherical coordinates $(r,\theta,\phi)$ and the corresponding spherical velocity components ($V_r, V_\theta,
V_\phi$) using the following convention\footnote{Note: the $\tan^{-1}(y/x)$ functions have been
  implemented in our {\tt Python} code using the 2-argument arctangent function {\tt numpy.arctan2(y,x)}, which takes into account the signs of $x$ and $y$.}:
\begin{equation}
r_\mathrm{gc} = \sqrt{X^2+Y^2+Z^2}
\end{equation}
\begin{equation}
\theta =\pi/2-\tan^{-1}(Z/\sqrt{X^2+Y^2})
\end{equation}
\begin{equation}
\phi = \tan^{-1}(Y/X)
\end{equation}
\begin{equation}
  V_r = (U\cos{\phi}+V\sin{\phi})\sin{\theta} + W\cos{\theta}
\end{equation}
\begin{equation}
  V_\theta = (U\cos{\phi} + V\sin{\phi})\cos{\theta} - W\sin{\theta}
\end{equation}
\begin{equation}
  V_\phi = U\sin{\phi} - V\cos{\phi}.
\end{equation}

{\it Gaia} parallaxes are available for a subset of the stars, but
only become competitive with the LAMOST distances for stars closer
than approximately 4 kpc. We therefore use LAMOST distances for all
our stars, as the cut $|Z|>5$ kpc (to remove disk and/or thick disk
stars) means matches to {\it Gaia} have larger relative errors in {\it
Gaia} parallaxes as compared to our relative photometrically
determined distance errors.

\section{Results}
\label{sec:results}

\subsection{Sample kinematics and the halo anisotropy}
\label{kinematics}

%--------------------------------------------------------------------------
\begin{figure*}
  \begin{tabular}{cc}
    \includegraphics[width=1\columnwidth]{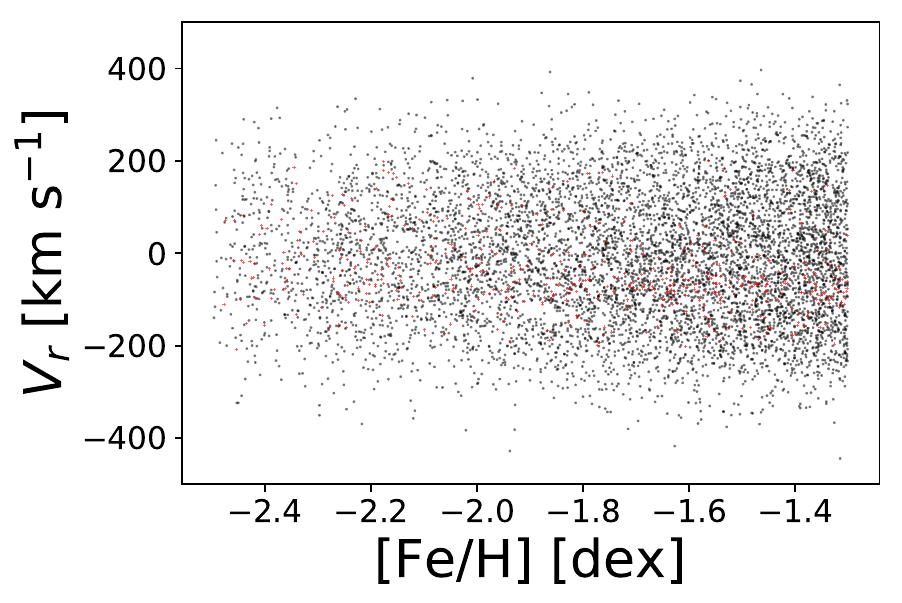}&
    \includegraphics[width=1\columnwidth]{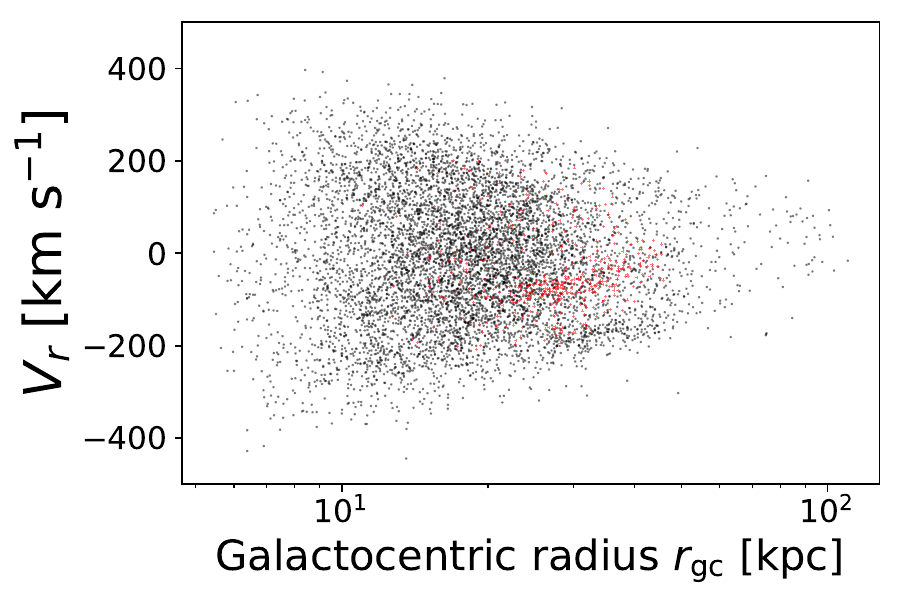}\\
    \includegraphics[width=1\columnwidth]{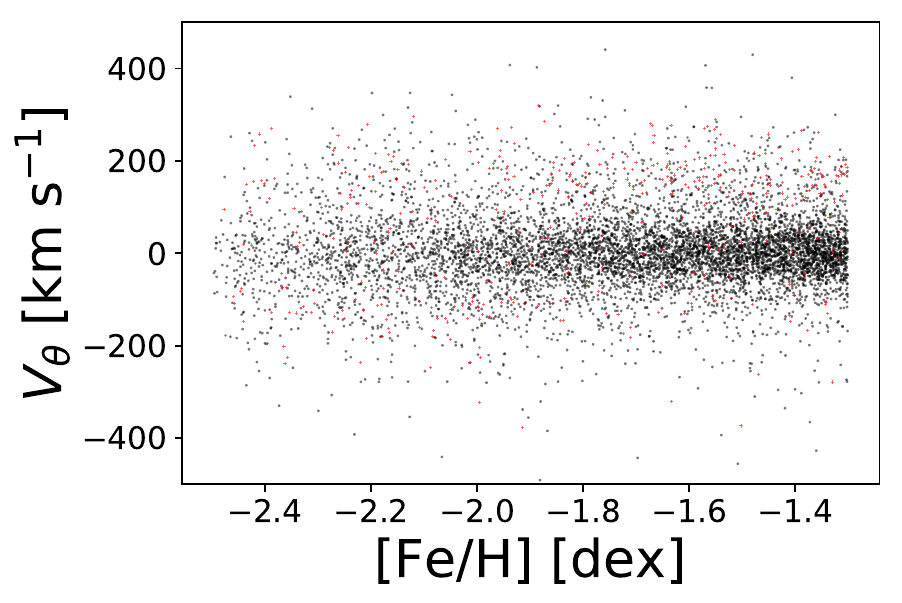}&
    \includegraphics[width=1\columnwidth]{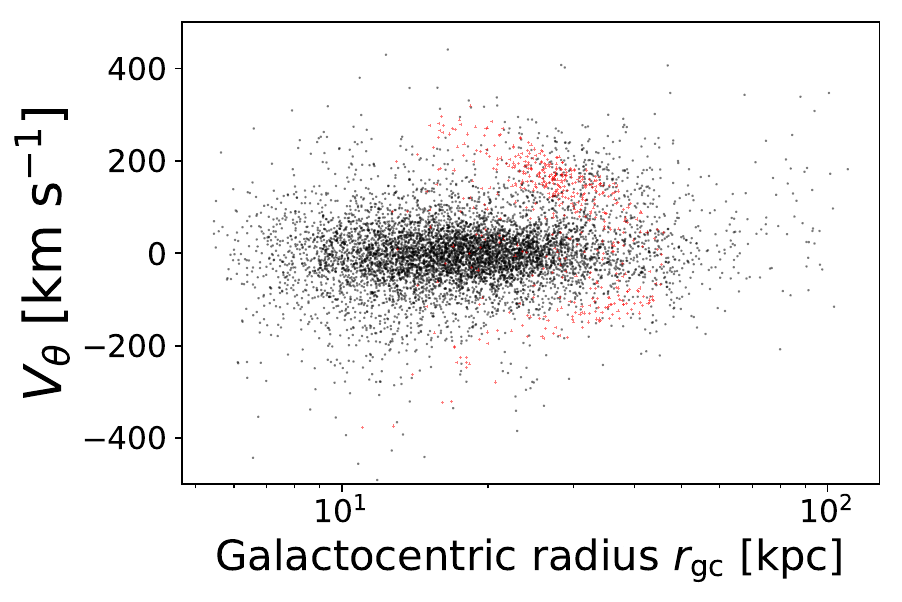}\\
    \includegraphics[width=1\columnwidth]{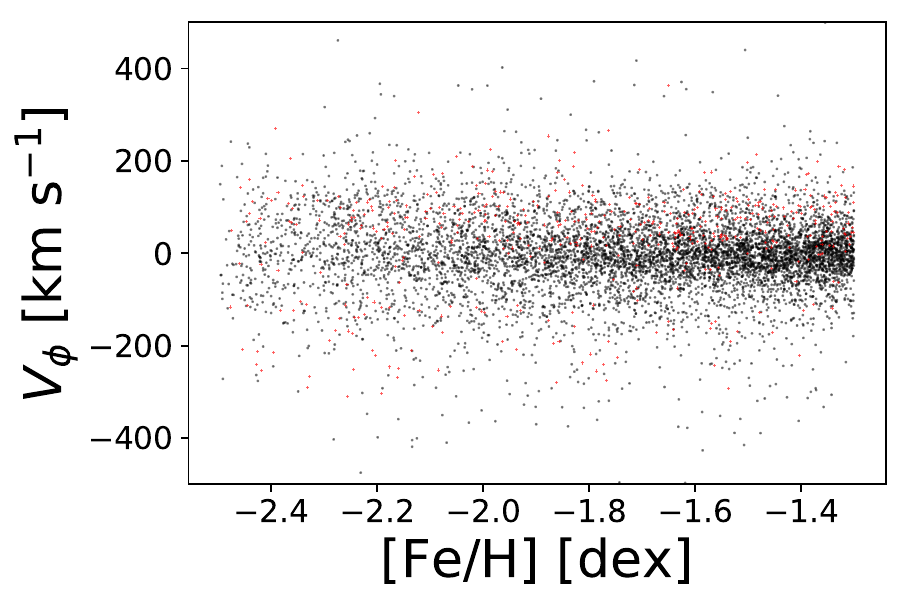}&
    \includegraphics[width=1\columnwidth]{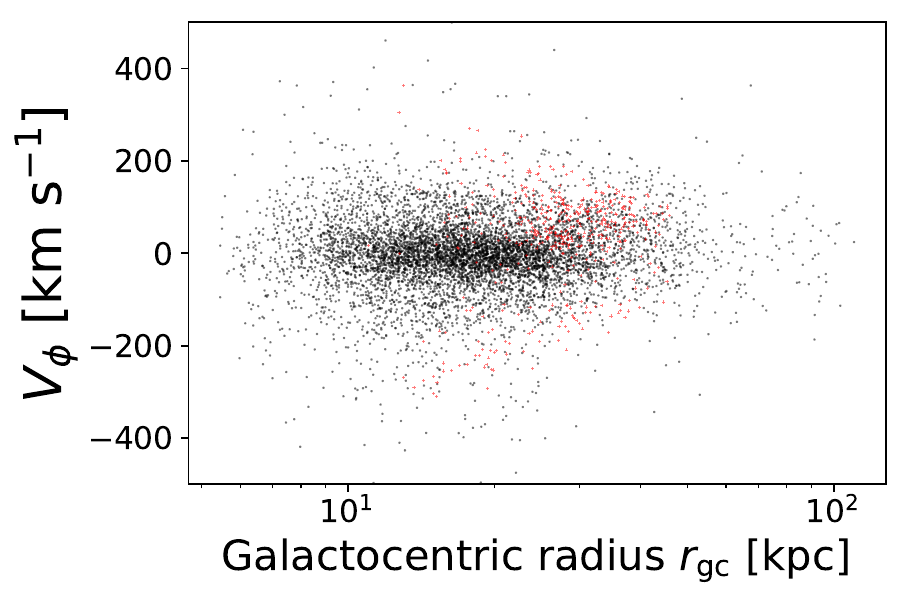}
  \end{tabular}
  \caption{The space velocities in spherical coordinates
    $(r,\theta,\phi)$ are shown for our sample of 7664 halo K
    giants (black dots and small red crosses) for which we have LAMOST line-of-sight velocities and
    metallicities, and {\it Gaia} proper motions. In the left 
    panels, the velocities are shown versus metallicity [Fe/H], and in
    the right panels versus Galactocentric distance $r_\mathrm{gc}$. We use the energy versus total angular momentum $E$-$L$ plane to flag an overdense region of stars as those predominantly belonging to Sagittarius (see Section \ref{sec:sag} and Fig. \ref{EL}). We propagate our $E$-$L$ selected Sagittarius-member stars in red in the left and right panels to view their characteristics in Galactocentric distance, velocity, and metallicity.
In the right
    panels in particular, several substructures in the kinematics are
    clear, including that due to the Sagittarius stream discussed in
    Section \ref{sec:sag}. The Galactocentric radial velocities
    $V_r$ show a considerably broader distribution than either of the
    tangential velocity components $V_\theta$ and $V_\phi$: the
    orbital families are predominantly radial. This becomes distinctly
    less the case for the lowest metallicities probed (left panels,
    [Fe/H] $ < -2$), as discussed in Section \ref{betametals},
    although the orbits are still predominantly radial.} 
  \label{3-D velocities}
\end{figure*}

%--------------------------------------------------------------------------

In Fig. \ref{3-D velocities}, we show the 3-D velocities $V_r,
V_\theta, V_\phi$ for our halo sample as functions of metallicity
[Fe/H] (left panels) and Galactocentric radius $r_\mathrm{gc}$ (right
panels). We have very conservatively cut the sample to metallicities
[Fe/H] $< -1.3$, based on our estimated metallicity errors of 0.1 dex,
in order to probe the halo only (examination of a fuller data set of
LAMOST K giants, for metallicities up to Solar ([Fe/H] $ = 0$), shows
that the disk-to-halo transition occurs at [Fe/H] $ = -1$ for the
stars analyzed (for which $|Z|>5$ kpc)). The K-giant velocities are
clearly those of a nearly non-rotating, pressure supported population,
with no sign of a much faster rotating thick disk (or disk) at our
highest metallicities ([Fe/H] $ = -1.3$). In the right-hand panels, we
show the velocities as a function of Galactocentric distance
$r_\mathrm{gc}$. Although the bulk of the stars are within 30 kpc of
the Galactic Center, we have sufficient numbers of stars to probe
$\beta$ out to 100 kpc.  There is very clearly substructure on
different scales in the velocity distributions versus Galactocentric
distance. 
The most obvious substructure is due to the Sagittarius
stream \citep[e.g.][]{Ibata1994,Majewski2003,Belokurov2006ApJ.642L.137}, seen in $(V_r,
V_\theta)$ at $V_r \approx -180, -80$ km s$^{-1}$ and $V_\theta \approx
200$ km s$^{-1}$.
In Section \ref{sec:sag} and Fig. \ref{EL}, we introduce a simple selection in energy $E$ and total angular momentum $L$ space which we use to remove the strongest overdense region visible by eye in $E$-$L$ space. These stars we associate as likely Sagittarius members and mark them as small red crosses in Fig. \ref{3-D velocities}.

\citet{Carlin2016shards} have analyzed the extent to which substructures
can be isolated in K giants in LAMOST, by comparing to ensembles of
smooth, synthetic Milky Way models finding that at least 10\% of
the stars can be associated with streams and other debris. Structures
become prominent beyond a Galactocentric radius of approximately 20
kpc within an otherwise well-mixed halo, while beyond approximately 40
kpc, star numbers in the catalog are insufficient to draw conclusions
on the sub-structure fraction. The strongest substructure they find
is due to the Sagittarius stream.

We next compare velocity dispersions and $\beta$ for the total data
set, with and without Sagittarius stars included. The aim is to
examine the extent to which the most prominent of substructures,
Sagittarius, affects the determination of $\beta$. Structures seen in
space and velocity coordinates are considered likely ``un-relaxed''
and ``non-virial'' components within the Galactic halo. As has been
shown in detail in \citet{Loebman2018}, the anisotropy parameter is
sensitive to both the presence of satellites in the halo, and to the
passage of satellites through the underlying, smooth and kinematically
relaxed stellar halo.

%--------------------------------------------------------------------------
\begin{figure}
    \includegraphics[width=0.8\columnwidth]{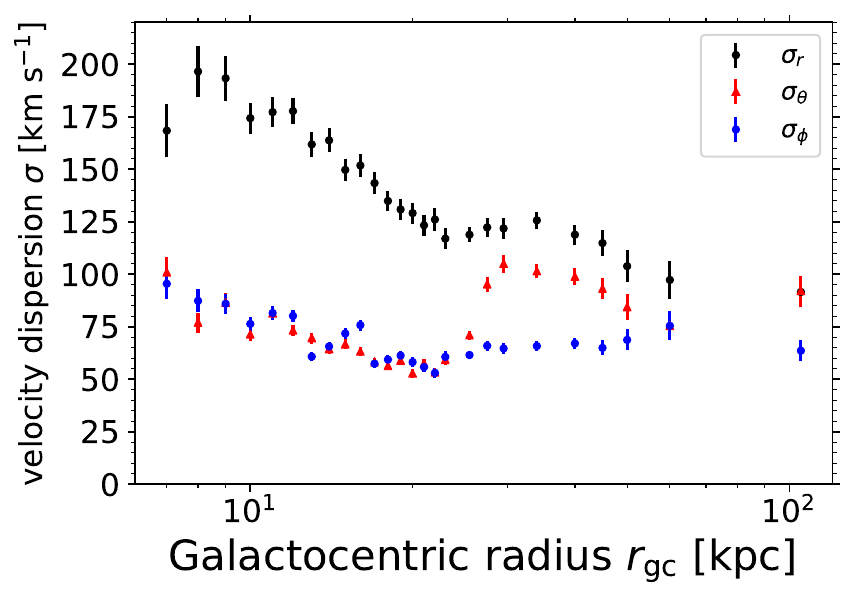}
    \includegraphics[width=0.8\columnwidth]{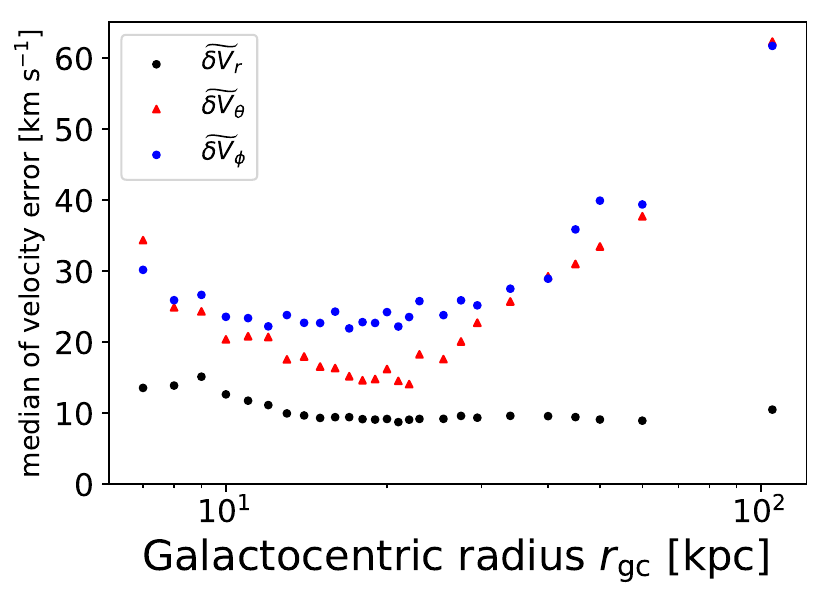}
    \caption{{\bf Upper panel:} three velocity dispersion components ($\sigma_r,
      \sigma_\theta, \sigma_\phi$)
computed from the total
      sample of 7664 halo K giants. Each marker represents the end of our selected radial bins. The first bin contains stars within $r_\mathrm{gc}=5-7$ kpc. Each consecutive bin is of width 1 kpc up to the bin ending at $r_\mathrm{gc}=30$ kpc, after which our bin sizes progressively increase. The remaining bins end 
at $r_\mathrm{gc}=$ 
32, 35, 40, 45, 50, 60, and 105 kpc.
Error bars are estimated from the Poissonian sampling in each bin.
{\bf Lower panel:} median errors 
($\widetilde{\delta V_r},\widetilde{\delta V_\theta}, \widetilde{\delta V_\phi}$)
on the velocities as
a function of Galactocentric radius, for the same bins as in the upper
panel.  The error estimates for each star are propagated from the
LAMOST distance errors, LAMOST line-of-sight velocity errors, and two
proper motion error estimates from {\it Gaia} DR2.  }
  \label{sigmas_medianerrs}
\end{figure}
%--------------------------------------------------------------------------

In the upper panel of Fig. \ref{sigmas_medianerrs}, we show the
velocity dispersions $(\sigma_r, \sigma_\theta, \sigma_\phi)$ as
functions of Galactocentric radius $r_\mathrm{gc}$. These have been
estimated from the component velocities using {\tt
  ROBUST\_SIGMA}\footnote{{\tt ROBUST\_SIGMA} \citep{Freudenreich1990}
  is a routine from IDL ASTROLIB \citep{Landsman1993}. We use a port
  to {\tt Python}.}, which is
designed particularly to reduce the effects of outliers in otherwise
relatively Gaussian distributions. We have subtracted in quadrature
the median of the velocity errors 
($\widetilde{\delta V_r},\widetilde{\delta V_\theta}, \widetilde{\delta V_\phi}$)
of the stars in each bin. These
errors are as propagated from the (1) errors in the LAMOST distances
of the stars (16\%), (2) the LAMOST line-of-sight radial velocity
errors (generally in the range 5 to 25 km s$^{-1}$), (3) and the two
proper motion error estimates from {\it Gaia} DR2.  For most bins the
corrections are quite small, but do grow with Galactocentric distance
as the stars become fainter, and the proper motions smaller, as seen
in the lower panel of Fig. \ref{sigmas_medianerrs}. In the most
distant bins the corrections are about the same magnitude as the
velocity dispersion we are aiming to measure. For less than 
1\% of the stars, velocities in excess of 500 km s$^{-1}$ were
obtained. These are very likely to be spurious rather than true high
velocity stars, and were removed from the final sample (such high
velocities are clipped by {\tt ROBUST\_SIGMA} in the analysis in any
case). Additionally, for approximately 1\% of the stars,
spuriously high errors were found for the $V_r$, $V_\theta$ and/or
$V_\phi$ velocities. Such stars were removed from the sample by
requiring that the error on the Galactocentric radial velocity be
$<100$ km s$^{-1}$, and the error on the tangential velocities be
$<150$ km s$^{-1}$. Tests with or without these criteria showed that
these cuts barely affect $\beta$, but we include them for
completeness.
  
The upper panel of Fig. \ref{sigmas_medianerrs} shows $\sigma_r$ and
the two tangential velocity dispersions $\sigma_\theta$ and
$\sigma_\phi$ as functions of Galactocentric radius for our complete
sample of halo K giants. It is clear that both tangential dispersions
are substantially less than the radial velocity dispersion $\sigma_r$
at all radii out to almost 100 kpc. The orbits are clearly strongly
radial throughout the halo.

%--------------------------------------------------------------------------
\begin{figure}
    \includegraphics[width=1\columnwidth]{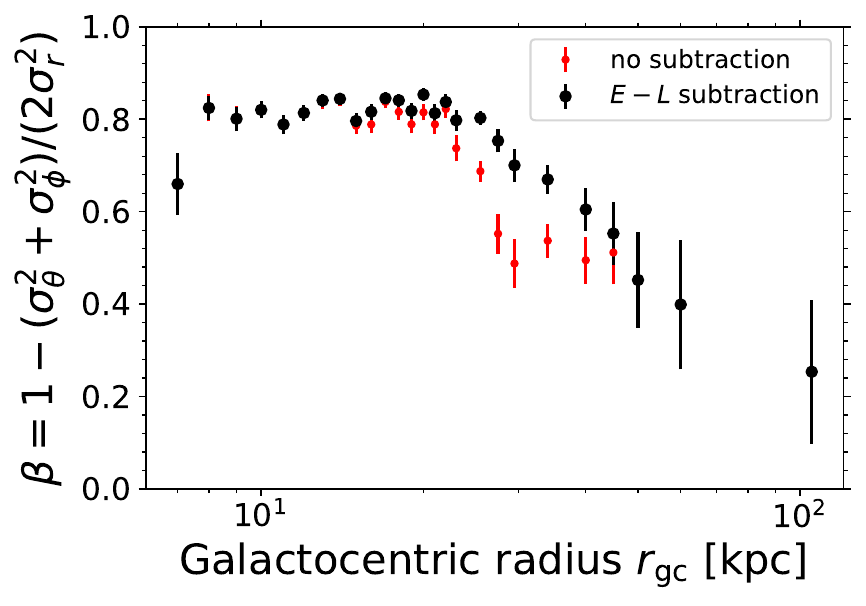}
  \caption{Anisotropy parameter $\beta$ as a function of
    Galactocentric radius $r_\mathrm{gc}$ for the entire sample of
    7664 halo K giants (red dots) and for the sample with
    Sagittarius flagged and removed (remaining sample of 7182 K giants, black dots), to highlight the
    effect of this substructure in the sample. The error bars on
    $\beta$ are propagated through from the errors in measuring
    $\sigma_r, \sigma_\theta$ and $\sigma_\phi$ (cf. Fig.
    \ref{sigmas_medianerrs}, upper panel). We
    see a dip from 25 to 40 kpc of the red-dot $\beta$ profile, due to the effects of the
    Sagittarius stream (we select the stream stars in Fig. \ref{EL}).
    Removing
    these stars takes a strongly tangential orbital family out from
    the sample, making $\beta$ more radial by $\sim 0.2$. 
}
  \label{rgc-beta}
\end{figure}
%--------------------------------------------------------------------------

%--------------------------------------------------------------------------
\begin{figure}
  \includegraphics[width=1\columnwidth]{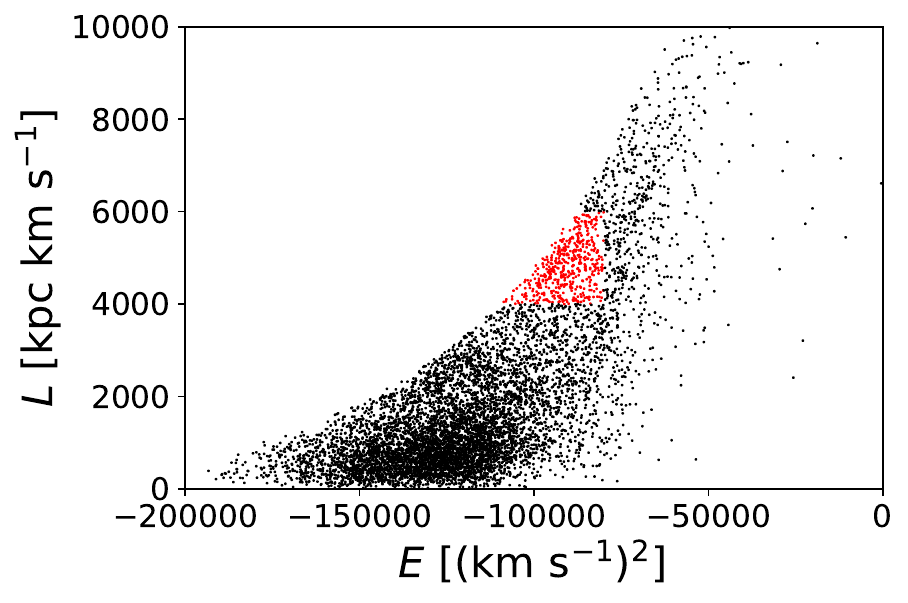}
  \caption{Energy $E$ versus total angular momentum $L$ for the full
    sample of 7664 halo K giants (black and red dots). Energies for
    the stars have been estimated using a Milky Way-like potential
    consisting of an NFW dark halo and a double exponential disk using
    {\tt galpy} \citep{Bovy2015}. 
We flag the overdense region of stars marked in red in this $E$-$L$ plane as predominantly in
    the Sagittarius stream.
}
  \label{EL}
\end{figure}
%--------------------------------------------------------------------------

We highlight this in Fig. \ref{rgc-beta}, where we show the orbital
anisotropy parameter $\beta$ versus Galactocentric radius
$r_\mathrm{gc}$ for the entire sample (i.e. without Sagittarius stars removed,
red dots). The halo orbits are strongly radial, with $\beta$
lying mainly in the range 0.6 to 0.8. It is remarkable how steady
$\beta$ is with Galactocentric radius $r_\mathrm{gc}$, out to about 25
kpc, despite the considerable change in the underlying $\sigma_r$ and
the two tangential velocity dispersions as functions of
$r_\mathrm{gc}$. Simulated stellar halos typically show a steady or very slowly rising $\beta$ profile for $r_\mathrm{gc}>5$ kpc 
(see e.g. \citet{Diemand2005}, \citet{Abadi2006},
\citet{Rashkov2013}), which is consistent with what we see out to
about 25 kpc in Fig. \ref{rgc-beta}. \cite{Loebman2018} have analyzed
the simulated stellar halos of \cite{Bullock2005},
\cite{Christensen2012}, and \cite{Stinson2013}, finding that (unless
the galaxies are in a state in which major merging is still taking
place) $\beta$ remains high (in the range 0.6 to 0.8) out to the
limits of their analysis (70 kpc). This is quite different to what we
see in Fig. \ref{rgc-beta}, where there is decline from $\beta \approx
0.8$ to $\beta \approx 0.2$ out to 100 kpc, at the limits of our
sample. 
To the best of our knowledge, no such behavior has been
    seen in simulated stellar halos.
It will be very interesting to examine to what extent this is
inconsistent with the halo stars seen in stellar halo simulations at
these large distances, work we leave for the future.

We have validated our analysis codes for determining $\beta$ as a
function of Galactocentric radius via mock data sets. The mocks are
based on the halo K-giant sample as follows: we set up a range of
models for $\beta(r_\mathrm{gc})$ as a function of Galactocentric radius $r_\mathrm{gc}$, 
and assign velocities $V_r, V_\theta, V_\phi$
to each star in the sample based on its Galactocentric distance and
the error estimates on $V_r, V_\theta, V_\phi$ propagated through from
the LAMOST and {\it Gaia} error estimates (in distance, line-of-sight velocities, and proper motions). Having imposed a particular functional form
for $\beta$ on the data set stars, we test our ability to recover it
using the methods applied to the actual data. We tested a range of
models for $\beta$: isothermal ($\beta = 0$), strongly 
radially 
anisotropic ($\beta = 0.75$), strongly tangentially
anisotropic ($\beta = -1$), and models for which $\beta$ is a function of $r_\mathrm{gc}$
\citep[e.g.][]{Sommer-Larsen1994}. 
We find that even with
the quite large errors that develop in the tangential velocities at
large distance (of order the velocity dispersion of the stars) seen in
Fig. \ref{sigmas_medianerrs}, we are able to recover $\beta$
correctly. 
These tests confirm that the dominant source of uncertainty
is Poisson statistics in the bins for our particular sample of halo
giants. We note that we provide no extensive test of anisotropy models with narrow features, 
such as the sharp dips in $\beta$ as a function of Galactocentric radius seen 
by \citet{Kafle2012}. Such features would be poorly resolved in 
this study of halo K giants because of distance errors, as already
pointed out by \citet{Hattori2017} and \citet{Loebman2018}, and the existence of 
such features is difficult to test with the current data set.

\subsection{Effects of substructure}
\label{sec:sag}

Simulations of stellar halos in Milky Way type galaxies show that the
anisotropy $\beta$ with Galactocentric radius is sensitive to the
presence of satellites and streams within the overall halo population,
as might be expected, but also to the passage of satellites affecting
the orbital parameters of the underlying, smooth, and kinematically
relaxed stellar halo \citep{Loebman2018}.

We do not make detailed analysis of such effects in this paper, but
restrict ourselves to a brief examination of Sagittarius (the most
prominent substructure in the velocity plots). There are a number of
ways in which its stars could be removed from the sample in order to
examine how they are currently affecting the $\beta$ profile: we have
chosen the simplest expedient of tagging Sagittarius in energy $E$
versus total angular momentum $L$.
We achieve this by equating Sagittarius with the most prominent overdensity in the $E$-$L$ plane. 
We note that $E$ versus the
vertical angular momentum $L_Z$ would be a better choice for isolating
Sagittarius stars over e.g. many orbits in an axisymmetric potential
as both these quantities are conserved (whereas $L$ may only be
approximately conserved), but that is not our aim here.

The full sample is shown in the $E$-$L$ space in Fig. \ref{EL}, where
we have adopted, again for simplicity, a Milky Way-like potential
$\Phi(R,Z)$ in the {\tt galpy} package \citep{Bovy2015} composed of
an NFW dark halo and a double exponential disk, to order the stars by
an energy, $E$:
\begin{equation}
  E = \Phi(R,Z) + V_{\mathrm{tot}}^2 / 2,
\end{equation}
where $R=\sqrt{X^2+Y^2}$ and $V$ is the total velocity of the star
($V_{\mathrm{tot}}^2 = U^2 + V^2 + W^2 = V_r^2 + V_\theta^2 +
V_\phi^2)$. The total angular momentum $L$ is
\begin{equation}
  L = r_\mathrm{gc} V_{\mathrm{tan}},
\end{equation}
where $V_{\mathrm {tan}}$ is the total tangential velocity
$(V_{\mathrm{tan}}^2 = V_\theta^2 + V_\phi^2)$.

To the eye, the $E$-$L$ plane shows 
    two major substructures in addition to the
    bulk of the halo stars seen at the bottom of the plot.
    Kinematic substructures appear prominently in $E$-$L$ space, due to common
    energy and angular momentum.
The most prominent overdensity in the $E$-$L$ plane is
found at
$E = -90000$ (km s$^{-1}$)$^2$ and $L = 5000$ kpc km s$^{-1}$. We equate this overdensity as being predominantly composed of Sagittarius stars and select a region surrounding the overdensity (in $E$-$L$ space) to cut from our sample as a simple method to test the effects on $\beta$ due to Sagittarius.
We define an arbitrary box in $E$-$L$ space with appropriate size to enclose the overdensity we wish to remove. We define limits of the region in which to remove the stars as
$E < -80000$ (km s$^{-1}$)$^2$ and $4000 < L < 6000$ kpc km s$^{-1}$
(marked in red in Fig. \ref{EL}) and recompute the $\beta(r)$ profile.
The stars separate out well in
this
space because of Sagittarius' high latitude, tangential orbit. 
We propagate our selection from the $E$-$L$ plane to the Galactocentric distance, velocity, and metallicity planes shown in Fig. \ref{3-D velocities} (small red crosses). Our selected stars from the $E$-$L$ space well-cover the most prominent features of Sagittarius seen in the $r_\mathrm{gc}$-$V_\theta$ space and $r_\mathrm{gc}$-$V_r$ space ($(V_r,V_\theta)\approx(-180,200)$ and $\approx(-80,200)$ km s$^{-1}$). That stars which we select with our simple cut in $E$-$L$ space also appear in the same region as Sagittarius in the Galactocentric distance and velocity spaces gives us confidence that we can use our simple $E$-$L$ as an initial exploration of the effects of Sagittarius on $\beta$.
After using the $E$-$L$ criteria to remove the stars that we associate with Sagittarius, we show the 
resulting profile in Fig. \ref{rgc-beta} (black dots).

The $\beta(r)$ profile with Sagittarius still in the sample is shown
by the red dots, and the profile with Sagittarius removed is
shown by the black dots. In the region where Sagittarius contributes
most strongly to the anisotropy profile, from 25 to about 40 kpc,
$\beta$ becomes more radial but only mildly so, rising by one to two
tenths. This is in line with expectation, as Sagittarius is on a high
latitude, tangential orbit, and biases $\beta$ to appear more
tangential (lower values of $\beta$). We note that outside the regions
in Galactocentric radius where Sagittarius is having an impact,
$\beta$ is barely affected.

This brief look at the most prominent substructure in the sample shows
that even a quite massive satellite passing through the underlying stellar
halo can have modest effects on $\beta$ in the sense that the overall stellar halo orbits
remain very radial. It is interesting that the break in the
$\beta$ profile, from a remarkably steady value at $\beta \approx 0.8$
from the Sun to 20 kpc, occurs where Sagittarius starts to make its
presence felt. This behavior, in which $\beta$ has breaks or dips as a
function of $r_\mathrm{gc}$, is similar to that seen in simulations of
the formation and evolution of the halo component in
$N$-body/hydrodynamical simulations analyzed in detail by
\citet{Loebman2018}. A simple view is that the break is from a quite
well mixed inner halo within approximately 25 kpc, where the orbital
families and $\beta$ are in (currently) a steady state to regions
where substructure is affecting $\beta$ via satellite infall and
streams. Clearly the effects of substructure are a major topic and we
leave this to future work, but point out that even the most prominent
substructure in our sample only affects $\beta$ modestly. It is
unlikely that the remaining, much weaker, substructures seen in the
velocity plots will affect the conclusion that the halo K giants in
our sample have strongly radial orbits.

\subsection{Effects of distance bias}
\label{sec:distance-beta}

\begin{figure}
  \includegraphics[width=1\columnwidth]{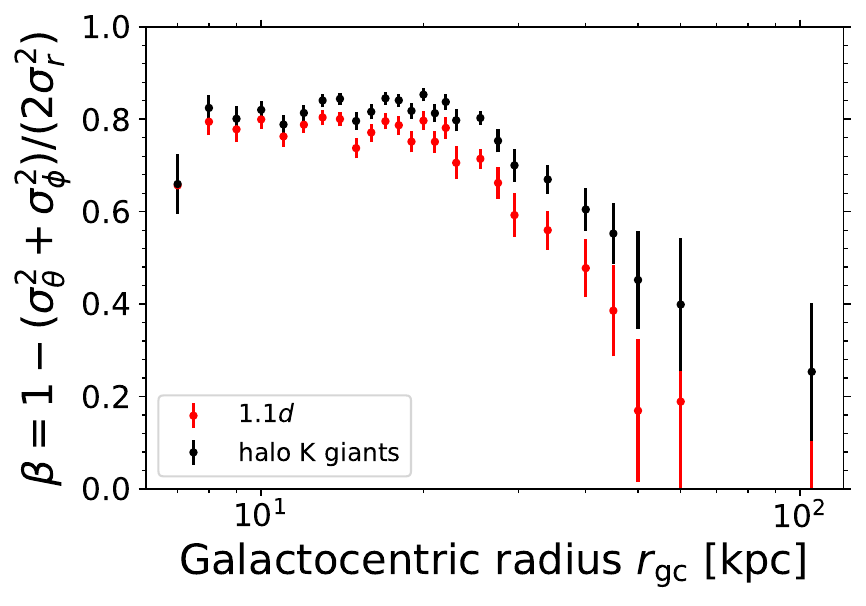}
  \caption{The effect on the estimates of the anisotropy parameter
    $\beta$ if we adopt a systematic correction to our distance scale
    for halo K giants of 10\%, to bring the giants into agreement with
    the distance scale for nearby halo K giants due to
    \citet{Bailer-Jones2018}. Black symbols show $\beta$ for the
    \citet{Xue2014} distance scale, and red symbols show $\beta$ for
    this distance scale increased by 10\%.  
}
  \label{fig:xue_cbj_beta}
\end{figure}

The adopted distance scale for our K giants was
    cross-checked in Section \ref{lamostkgiants}, where we compared
    our halo K-giant distance scale to that of
    \citet{Bailer-Jones2018}, for halo K giants within approximately 4
    kpc of the Sun. This test showed evidence for our distances being
    closer on average than the \citet{Bailer-Jones2018} distances by
    approximately 10\%
    (Fig. \ref{fig:distance-compare}). \citet{Bailer-Jones2018} use a
    model of a ``lengthscale'' at any given $(l,b)$ position, to
    derive improved distance estimates of stars compared to simply
    inverting the parallax. The modeling is validated using stars in
    open clusters, i.e. on metal rich stars. The \citet{Xue2014}
    distances have been calibrated using K giants in globular
    clusters, stars very similar to our halo K giants. A systematic
    distance correction for our sample stars will have the
    effect of increasing the estimated tangential velocity dispersions
    (which are primarily sensitive to the distance scale via proper
    motions), while leaving the Galactocentric radial velocity
    dispersions relatively unaffected (as these are typically
    dominated by line-of-sight velocities). Consequently we expect
    such a distance correction to reduce the estimates of the
    anisotropy $\beta$ slightly. We show this in
    Fig. \ref{fig:xue_cbj_beta} using our halo K-giant sample after removing Sagittarius.  The anisotropy $\beta$ decreases very
    slightly in the inner halo ($r_\mathrm{gc}<30$ kpc) as a result of
    this correction although it decreases somewhat more in the outer
    halo. This shows that the main result of the paper, that the bulk
    of the halo K giants are on highly radial orbits, is very robust
    to systematic error in the distance scale of this order.

%--------------------------------------------------------------------------
%--------------------------------------------------------------------------

%--------------------------------------------------------------------------
\begin{figure*}
  \begin{tabular}{cc}
\includegraphics[width=1\columnwidth]{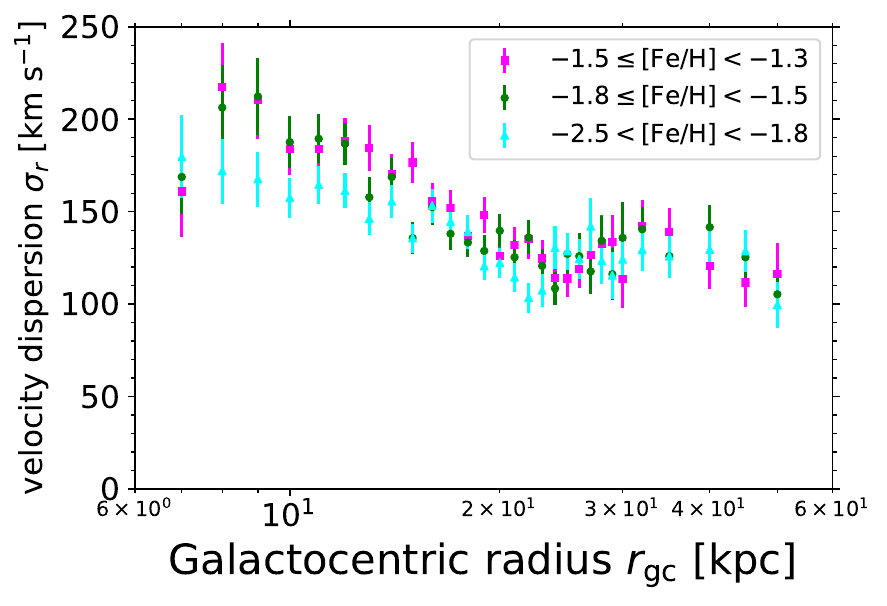}&
\includegraphics[width=1\columnwidth]{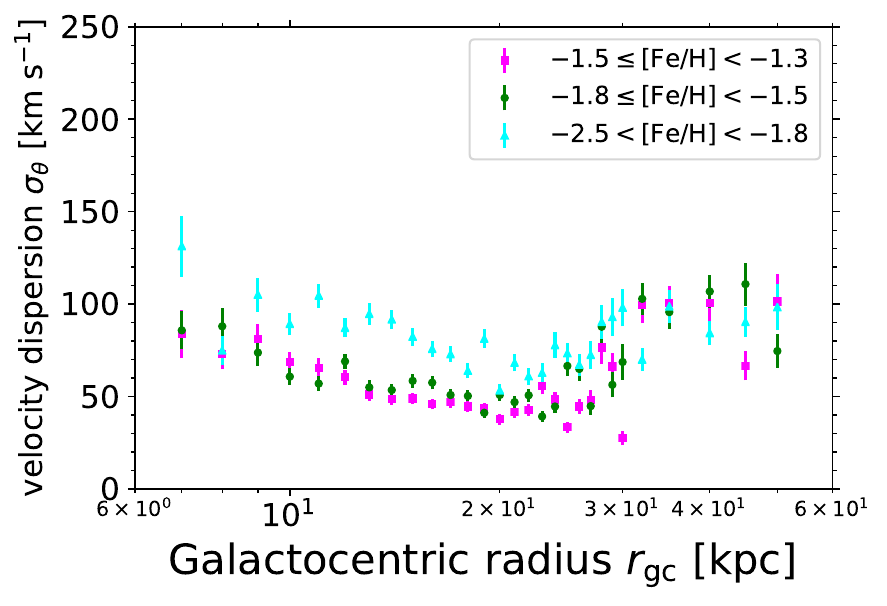}\\
\includegraphics[width=1\columnwidth]{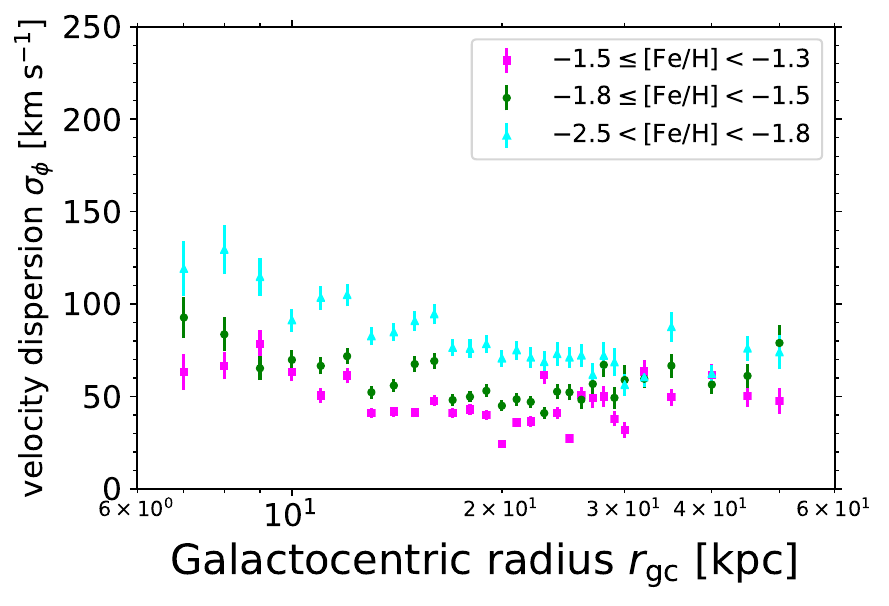}&
    \includegraphics[width=1\columnwidth]{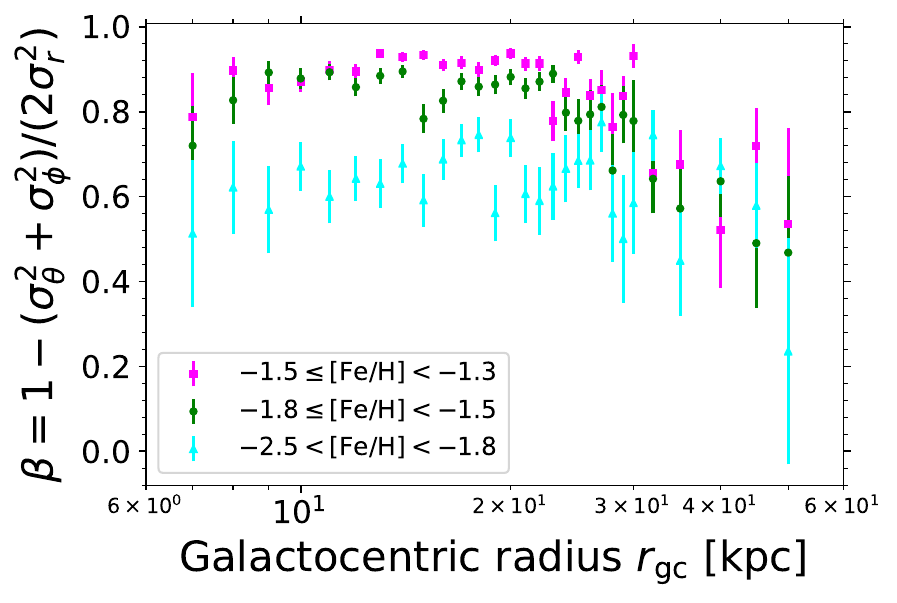}
  \end{tabular}
  \caption{Profiles in Galactocentric radius $r_\mathrm{gc}$, of the
    three velocity dispersion components $\sigma_r, \sigma_\theta, \sigma_\phi$, 
and anisotropy parameter $\beta$, for three
    metallicity bins, using our sample of halo K giants after efforts
    to remove Sagittarius stars as in Fig. \ref{rgc-beta}.  The top left, top right, and lower left panels show the three dispersion profiles, $\sigma_r, \sigma_\theta, \sigma_\phi$, respectively, divided into three metallicity bins. The lower right panel shows the anisotropy profile $\beta$.
The two
    more metal rich bins share a common amplitude of $\beta\approx0.9$
    with $r_\mathrm{gc}<25$ kpc and gradually declining value of
    $\beta$ at larger radii.  The anisotropy profile is strikingly
    different for stars with [Fe/H]$<-1.8$, where the amplitude of
    $\beta$ has dropped to $\approx 0.6$ and remains relatively
    constant out to large radii; such stars are still on quite radial
    orbits, but substantially less so than the rest of the
    sample. This change in the underlying velocities used to form
    $\beta$ can also be seen in the left-hand panels of Fig. \ref{3-D
      velocities}.  We note that substructure,
    particularly beyond
    $\approx$ 25 kpc, still is affecting $\beta$ and the behavior
    should be regarded with care until careful structure identification and removal is performed . }
  \label{betafeh}
\end{figure*}
%--------------------------------------------------------------------------

\subsection{Metallicity dependence of the anisotropy}\label{betametals}

We have examined the behavior of $\beta(r)$ as a function of the
metallicity of the stars. After removing Sagittarius stream stars,
selecting stars over the metallicity range $-2.5 < $ [Fe/H] $ < -1.3$,
and dividing them into three roughly equally populated bins in this
range, we have plotted ($\sigma_r, \sigma_\theta, \sigma_\phi$) and
$\beta$ versus $r_\mathrm{gc}$ for each subset, as shown in Fig.
\ref{betafeh}.  Interestingly, for the metal poor bin ($-2.5 <$ [Fe/H]
$< -1.8$), we find a significantly lower $\beta$ at 0.6, compared to
0.9 for the rest of the sample.  Beyond 25 kpc, the two more metal
rich bins steadily decline in anisotropy whereas the more metal poor
bin remains relatively constant; but we note that substructure clearly
affects $\beta$ and any trends should be treated with caution.

A picture has emerged in recent years of a two-component halo which
differs in spatial distributions of metallicity, age, and kinematics
\citep[e.g.][]{Carollo2007,Deason2011.411}.
The same formation and evolution processes responsible for the 
dichotomy found in the stellar halo are
also likely to affect the anisotropy of the orbits.
Indeed, \citet{Hattori2013} and \citet{Kafle2013}, in studies of halo BHB stars within a few tens of kpc, have noted
that anisotropy is a function of metallicity. In both studies, the
halo stars are subdivided into two metallicity bins at [Fe/H] $=-2$, and the difference in
the kinematics (i.e. anisotropy and the bulk rotation) of the two
sub-samples is argued to support the two-component picture for the
halo. They find that the lower metallicity stars have rounder orbits
--- although the anisotropy obtained is significantly lower than what
we find for our metal poor stars (we find $\beta \approx 0.6$ whereas
they find much rounder orbits, with $\beta \approx 0$ to $-1$).

It is very instructive to compare our sample to
\citet{Belokurov2018}, who have analyzed the anisotropy
properties of a sample of main sequence halo stars, for which, as
here, full kinematical and metallicity data are available. Their
sample have been selected from SDSS and {\it Gaia} DR1, and are within
$\approx 10$ kpc of the Sun. They find similar results for nearby
stars to what we have found in the distant halo. The orbits of the
bulk of the nearby halo stars are highly radial $\beta \approx 0.9$,
with $\beta$ declining to more mildly radial orbits (0.2 $< \beta <$
0.4) down to the lowest metallicities probed ([Fe/H]$ \approx
-3$). This decline in $\beta$ with metallicity is similar to what we
find for the halo K giants out to about 25 kpc. At a metallicity of
[Fe/H] $ \approx -2$, their Fig. 4 shows $\beta \approx 0.5$, quite
similar to what we find ($\beta = 0.6$) in our most metal poor bin
($-2.5 <$ [Fe/H] $< -1.8$). Beyond about 25 kpc, our sample becomes
much more obviously affected by structure, and comparison with
\citet{Belokurov2018} is not straightforward, but within 25 kpc
the two samples do appear to be sampling a very similar component of
the Galactic halo in terms of the anisotropy dependence on [Fe/H].

Fig. \ref{betafeh} shows a dip in $\beta$ at a Galactocentric radius
of approximately 15 kpc for the stars in the metallicity range $-1.8 <
$ [Fe/H] $ < -1.5$. This is reminiscent of the dips seen in the halo
simulations analyzed by \citet{Loebman2018}, due to a disrupted
satellite passage through a well-mixed background halo. Such dips
would be short-lived and have metallicities restricted to those of the
dissolving satellite, as opposed to long-lived dips which involve
stars in the in-situ halo, and thus no restriction to a particular
range of metallicity. It will be interesting to see if this picture
can be confirmed as part of an ongoing search for and analysis of
streams amongst the LAMOST halo K giants.

\section{Conclusions} \label{sec:conclusion}

The combination of distances and line-of-sight velocities from LAMOST,
and proper motions from {\it Gaia}, demonstrate the power that 6-D
surveys bring to Galactic structure and evolution studies. We have
measured the anisotropy parameter $\beta$ for K-giant stars in the
Galactic halo, with metallicities [Fe/H] $< -1.3$, over a range of
Galactocentric radii from 5 to over 100 kpc. We find that from the
Solar position to about 25 kpc, the orbits are highly radial ($\beta
\approx 0.8$) and then generally decline to near isothermallity (with
$\beta = 0.3$) to the limit of current measurements ($\sim $100 kpc).
The anisotropy profile in the region $r_\mathrm{gc} > 25$ kpc shows
very clear signs of being affected by substructure, in line with
expectations from $N$-body/hydrodynamical simulations of stellar halo
formation and evolution around Milky Way-type galaxies
\citep{Loebman2018}. Within the better relaxed region ($r_\mathrm{gc}
< 25$ kpc), our halo K giants share striking similarities with the
halo main sequence star sample of \citet{Belokurov2018}, who find
very high anisotropy of $\beta \approx 0.9$ for the bulk of their
stars, and a similar metallicity dependence, with orbits becoming more
mildly radial with decreasing metallicity. We have
    compared our distance scale for nearby halo K giants (within 4
    kpc) with that of \citet{Bailer-Jones2018} (based on {\it Gaia}
    parallaxes and priors for the stellar distribution in the Galaxy),
    finding that the scales differ by of order 10 \% (in the sense
    that our K giants have distances 10\% closer than the estimates of
    \citet{Bailer-Jones2018}). The adoption of a
    \citet{Bailer-Jones2018} distance scale reduces the anisotropy
    measure only slightly for our sample, leaving the overall
    conclusion of a highly radially anisotropic halo unchanged.

The end-of-mission {\it Gaia} proper motions will improve
substantially over DR2, particularly for the more distant stars, so
that we will be able to probe the anisotropy of the Galactic halo
significantly better in the regions where it appears to be declining.

The determination of the total mass profile of the Milky Way (i.e. the
enclosed mass as a function of Galactocentric radius) can be made
using the Jeans equation and the full 6-D data for our sample: in the
simplified case of a spherical Jeans equation, it is known that
estimation of the mass profile is biased, even if $\beta$ is known.
The mass profile, although still subject to some sample bias, in
particular due to non-virial effects, can be obtained using the 3-D
version of the Jeans equation. The velocity dispersion profiles
presented here will allow this, and we intend to measure the mass
profile in a forthcoming paper.

\acknowledgments

We thank Warren Brown, Monica Valluri, Emily Cunningham, and Huang Yang for useful discussions.
S.A.B. acknowledges support from the Postdoctoral Scholar's Fellowship
of LAMOST and the Chinese Academy of Sciences President's
International Fellowship Initiative Grant (no. 2016PE010) and the Aliyun Fellowship.  The
research presented here is partially supported by the 973 Program of
China under grant no. 2014CB845700, by the National Natural Science
Foundation of China under grant nos. 11773052, 11333003, 11761131016,
and by a China-Chile joint grant from CASSACA. J.S. acknowledges
support from a {\it Newton Advanced Fellowship} awarded by the Royal
Society and the Newton Fund, and from the CAS/SAFEA International
Partnership Program for Creative Research Teams. This work made use of
the facilities of the Center for High Performance Computing at
Shanghai Astronomical Observatory.  
X.-X.X. acknowledges support from the 
``Recruitment Program of Global Youth Experts'' of China and NSFC under grants 
11390371, 11873052, and 11890694.
C.F. acknowledges financial
support by the Beckwith Trust. 
C.L. acknowledges the National Key Basic Research Program of China 2014CB845700 and the NSFC under grant 11333003.
Guoshoujing Telescope (the Large Sky
Area Multi-Object Fiber Spectroscopic Telescope LAMOST) is a National
Major Scientific Project built by the Chinese Academy of
Sciences. Funding for the project has been provided by the National
Development and Reform Commission. LAMOST is operated and managed by
the National Astronomical Observatories, Chinese Academy of
Sciences. This work has made use of data from the European Space
Agency (ESA) mission {\it Gaia}, processed by the {\it Gaia} Data Processing and
Analysis Consortium (DPAC). Funding for the DPAC has been provided by
national institutions, in particular the institutions participating in
the {\it Gaia} Multilateral Agreement. This research has made use of NASA's
Astrophysics Data System Bibliographic Services.

\software{{\tt Astropy} \citep[v2.0.2;][]{astropy2013,astropy2018},
{\tt galpy} \citep[1.3.0;][]{Bovy2015}, {\tt matplotlib.pyplot} \citep[2.1.0;][]{Hunter2007}, NumPy \citep[v1.13.3;][]{Oliphant2006,Walt2011,Oliphant2015}, {\tt ROBUST\_SIGMA} \citep{Freudenreich1990}; \citep[{\tt ASTROLIB};][]{Landsman1993}, {\tt TOPCAT} \citep[v4.3-3;][]{Taylor2005} \explain{Per the new AAS software policy, \\ http://journals.aas.org/policy/software.html, we modify our AASTeX v6.2 manuscript to highlight the code we used (both cited and unmentioned in the current text) with the new {\tt \textbackslash software} command.}}

\listofchanges 

\end{document}